\documentclass[11pt,epsfig]{article}
\usepackage{epsfig}
\usepackage{amssymb,amsmath,latexsym}
\usepackage{natbib}
\title{{\bf Spectrophotometric Investigation of\\a Sample of Tidal Dwarf Galaxies}}
\author{Peter~M.~Weilbacher$^1$, Pierre-Alain~Duc$^2$\\
\vspace{0.1cm}\\
\normalsize $^1$Universit\"ats-Sternwarte G\"ottingen, Geismarlandstr.~11, 37083 G\"ottingen, Germany\\
\normalsize $^2$CNRS and CEA/DSM/DAPNIA/SAp, Saclay, 91191 Gif sur Yvette cedex, France\\
}
\date{}
\begin{document}
\maketitle
\def\bull{\vrule height .9ex width .8ex depth -.1ex}
\makeatletter
\def\ps@plain{\let\@mkboth\gobbletwo
\def\@oddhead{}\def\@oddfoot{\hfil\tiny
``Dwarf Galaxies and their Environment'';
International Conference in Bad Honnef, Germany, 23-27 January 2001}%
\def\@evenhead{}\let\@evenfoot\@oddfoot}
\makeatother

\begin{abstract}\noindent
  We define a Tidal Dwarf Galaxy (TDG) as a self-gravitating entity of
  dwarf-galaxy mass built from tidal material expelled during interactions.
  We summarize our findings on broad-band imaging and spectroscopy of
  a sample of Tidal Dwarf Galaxies candidates in a sequence of
  interacting systems.  Evidence for decoupled kinematics in the
  ionized gas have been found in several objects. This could indicate
  that they are bound galaxies and therefore genuine TDGs.  As a
  detailed example we analyze the system {\sf AM 1159-530}, where
  surprisingly high velocity gradients have been measured.
\end{abstract}

\section{Introduction}

In the past several authors have used the term ``Tidal Dwarf Galaxy''
to describe everything from faint knots with dubious association to
the interacting system and luminous HII regions to massive knots
embedded in tidal tails. We would now like to give a more 
restrictive\\[-10pt]

\begin{center}
  \fbox{
    \parbox[c]{0.8\linewidth}{
      \begin{center}
        {\LARGE Definition\\[5pt]}
         {\large We define a Tidal Dwarf Galaxy (\textbf{TDG}) as a self-gravitating entity of
           dwarf-galaxy mass built from tidal material expelled during interactions.}
      \end{center}
      }
    }
\end{center}
    
This kind of {\it true TDG} is difficult to prove observationally, so
that in most of the real cases one is restricted to observing {\it
  decoupled kinematics} (as a hint to dynamical stability) of a
condensation in or at the end of a tidal tail. Systems where this has
been accomplished up to now are \textsf{Arp 105}, \textsf{NGC 7252},
\textsf{NGC 5291}, and \textsf{Arp 245}
\citep[][resp.]{DBW+97,HGvG+94,DM98,DBS+00}.\\

From today's observational evidence and dynamical and photometric
modelling, a formation sequence for TDGs can be drafted: \\
An interaction between two giant galaxies occurs; at least one has to
be gas rich to account for the star-formation observed in TDGs.  Due
to gravitational forces and depending on the encounter geometry tidal
tails form out of the disk material. One or more condensations may
appear in the gaseous and/or stellar component of the tail.  Other
material is then pulled into this gravitational potential well.
Collapsing gas may then initiate a starburst on top of the old stellar
population inherited from the parent galaxy. After a long time tidal
features disappear, while the condensations may survive as ``normal''
dwarf galaxies.

\section{Imaging Results}

\citet{WDF+00} performed a photometric investigation of TDGs in 10
interacting systems using broad-band $B$,$V\!$,$R$ imaging. Using
two-color plots (see Weilbacher \& Fritze-v.Alvensleben, this volume)
probable TDGs were identified using their distinct color range (which
differs from that of background galaxies) in comparison with
photometric evolutionary synthesis models. Finally {\it 36 candidates
  for TDGs} were selected from about 100 knots in the tidal tails. It
was found that these TDG candidates are brighter than normal HII
regions in spiral galaxies, and are experiencing {\it strong
  starbursts}. Therefore a {\it strong fading} is expected, if/after
the current burst stops.

\section{Spectroscopic Sample Overview}

To be able to confirm the photometric results of our TDG candidate
selection we conducted spectroscopic observations of the sample of
interacting systems listed in Table~\ref{tab:sample}. We used the
EFOSC2 instrument on the ESO-3.6m telescope.  The table lists the
spectroscopic observing mode (multi-object or long-slit spectroscopy)
for each system. The last four columns give the number of
photometrically selected TDG candidates, the number of condensations
that we could detect (restrictions were brightness and positioning of
MOS slits), the number of objects with emission lines, and finally the
number of objects where kinematical signatures are visible on the 2D
spectrum.

\begin{table}[htbp]
  \begin{center}
    \begin{tabular}[h]{l | l | c c c c}
      \hline
      System name          & Observ. & \multicolumn{4}{c}{TDG-Candidates} \\
                           & mode    & total & detected & w.~emiss.~lines & kinematics vis. \\
      \hline
      \hline
      \textsf{AM 0529-565} & MOS     &     4 &        2 &                &                 \\
      \textsf{AM 0537-292} & 2 MOS   &     5 &        5 &              4 &               2 \\
      \textsf{AM 0547-244} & MOS     &     3 &        2 &              2 &             1(2)\\
      \textsf{AM 0547-474} & LS      &     2 &        2 &                &                 \\
      \textsf{AM 0607-444} & MOS     &     1 &        1 &              1 &                 \\
      \textsf{AM 0642-645} & LS      &       &          &                &                 \\
      \textsf{AM 0748-665} & MOS     &     3 &        1 &              1 &                 \\
      \textsf{AM 1054-325} & MOS+LS  &     6 &        5 &              5 &               2 \\
      \textsf{AM 1159-530} & MOS     &     1 &        1 &              1 &               1 \\
      \textsf{AM 1208-273} & MOS     &       &          &                &                 \\
      \textsf{AM 1237-364} & MOS     &     4 &        4 &              4 &               2 \\
      \textsf{AM 1324-431} & MOS     &     4 &        2 &              2 &              (1)\\
      \textsf{AM 1325-292} & LS      &     2 &        2 &             (2)&                 \\
      \textsf{AM 1353-272} & MOS     &    12 &        4 &              4 &                 \\
      \hline
    \end{tabular}
    \caption{Spectroscopic observations summary}
    \label{tab:sample}
  \end{center}
\end{table}

To illustrate how we will analyze the data of these systems, we
present one of our sample systems as an example.

\section{The system {\sf AM 1159-530}}

\begin{figure}[htbp]
  \begin{center}
    \epsfig{file=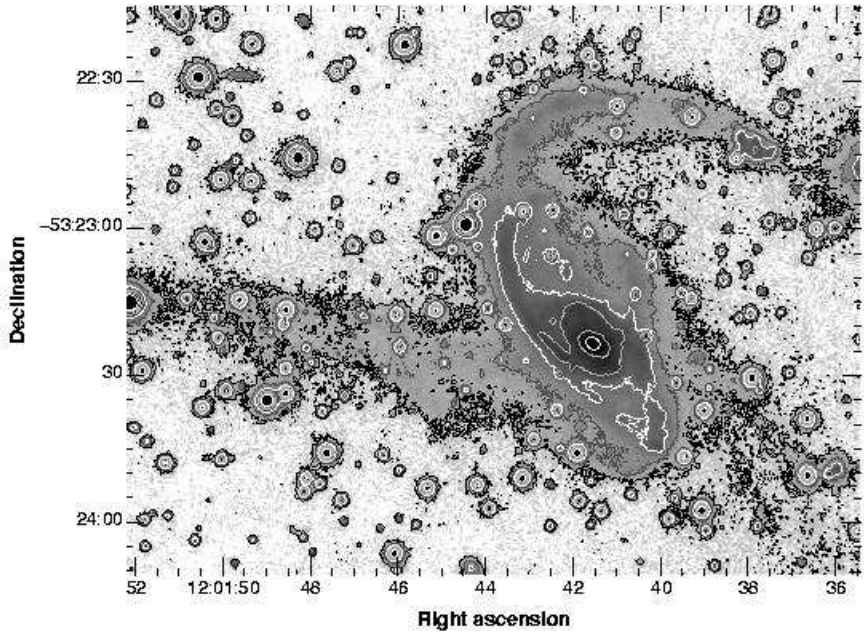,angle=-90,width=0.7\linewidth}
    \caption{The disturbed system {\sf AM 1159-530}. Surface brightness 
             contours from 20 to 25 mag/$\Box$'' are indicated. 
             The scale is 10'' = 3 kpc. }
    \label{fig:AM_1159-530}
  \end{center}
\end{figure}

This is a strongly disturbed system with two tidal tails but without a
clearly identified interacting companion, which might have caused this
disturbed appearance. The luminosity of the nucleus is $M_B = -18.4$
mag, the central receding velocity $V_\mathrm{nucl} = 4530\text{ km
  s}^{-1}$, corresponding to a distance of $D = 60$ Mpc when using a
Hubble constant of $H_0 = 75\text{ km s}^{-1}\text{ Mpc}^{-1}$. The
tidal tails of the system have projected lengths of 38 and 29 kpc
(western and northern tail, resp.).  In Fig.~\ref{fig:AM_1159-530} a
logarithmic intensity image of the system is shown.

At the end of the northern tail a bright ($M_B = -16.3$ mag)
condensation is visible, which we selected as a TDG candidate.  The
optical extent of this object amounts to 3.6 kpc $\times$ 2.1 kpc, and
it has a knotty appearance. It also has blue colors of $B-V = 0.18$
and $B-R = 0.30$ mag and hosts a HII region. Its spectrum exhibits a
weak continuum and strong emission lines. From the Balmer line fluxes
we derive a star formation rate of $\sim$1 M$_\odot$ yr$^{-1}$
\citep[from the calibration of][]{KTC94}, and using the Balmer and
oxygen lines we estimate an oxygen abundance of $12+\log(O/H) =
8.4\pm0.1$, shifting the object from the luminosity-metallicity
relation of normal dwarf galaxies into the region where most TDGs are
located (see contribution by P.-A. Duc in this volume).

\begin{figure}[htbp]
  \begin{center}
    \epsfig{file=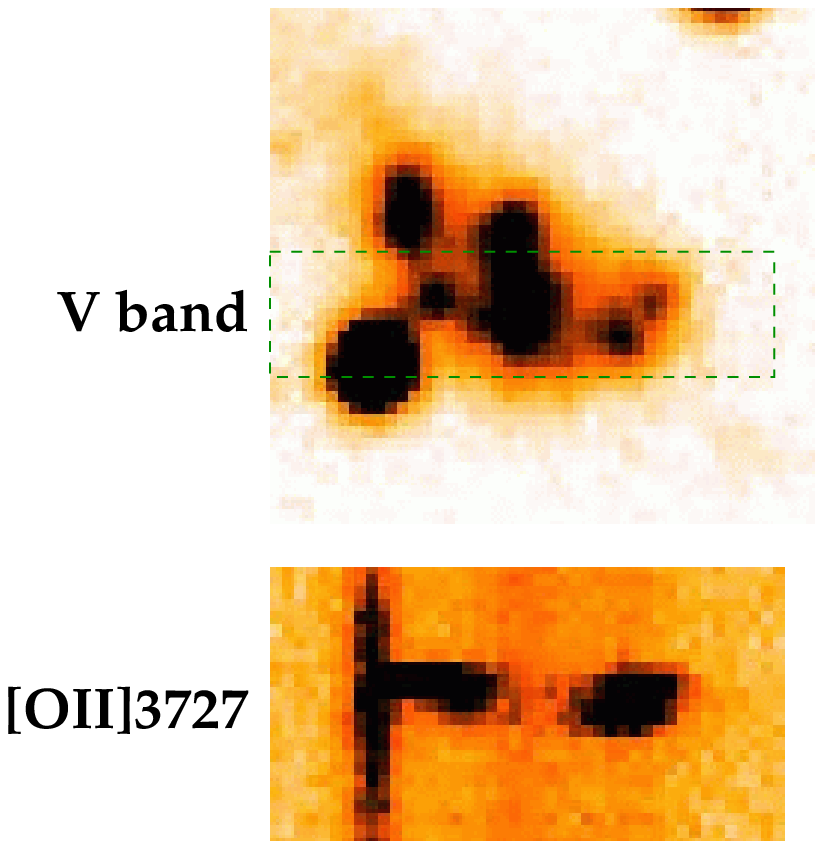,width=0.35\linewidth}
\psfull
    \epsfig{file=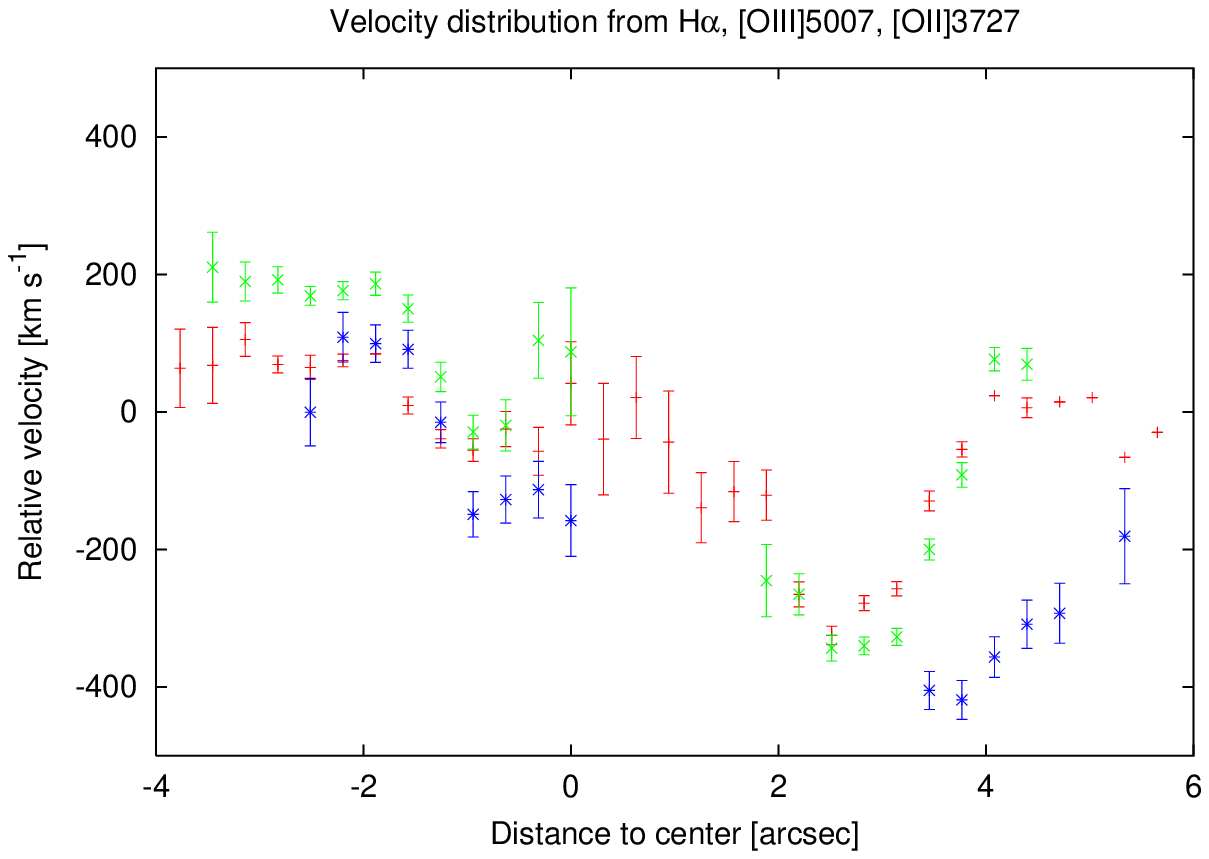,width=0.6\linewidth}
    \caption{{\bf (a)} V-band image of the TDG with the position of the slit
                        marked and the position-velocity distribution as 
                        extracted from the 2D spectrum from the region around
                        the [OII]3727 line. 
             {\bf (b)} Velocity distribution as derived from the three lines 
                        H$\alpha$, [OIII]5007, and [OII]3727.}
    \label{fig:rot}
  \end{center}
\end{figure}

The ionized gas in this object has a strong velocity gradient which is
clearly visible in the 2D spectra shown in Fig.~\ref{fig:rot}{\bf
  (a)}. The corresponding velocity curve is plotted in
Fig.~\ref{fig:rot}{\bf (b)}. We measure a peak to peak gradient as
large as $\sim$500 km s$^{-1}$ within a diameter of only 1.7 kpc. For
reference, the mean velocity of the tail itself varies by less than 80
km s$^{-1}$ from its base to its tip.  Pure streaming motions can
therefore only account for a few percent within the small region of
the TDG. Such a velocity gradient could, in principle,
trace the re-accretion of tidal material or be interpreted in terms of
rotation. If the latter were the case, one would derive from it a
virial mass $M_\mathrm{dyn}$ as large as $10^{10} M_\odot$.  This is
of course much higher than the stellar mass, which is estimated to be
$M_\mathrm{stars} = 10^{8} M_\odot$ from the optical luminosity. But
given the errors in both of these estimates and the fact that the
dynamical state of the object is largely unknown, we do not think that
dark matter is needed to fill the gap. The gas mass, which is usually
high in TDGs \citep[e.g.][]{DBW+97} but currently unknown in {\sf AM
  1159-530}, might account for the missing mass.

In any case, the condensation at the end of the northern tail of {\sf
  AM 1159-530} appears to be kinematically decoupled from the tidal
tail and thus is a {\it true} TDG.

\section{Conclusions}

We have found one more system with a bright TDG kinematically
decoupled from the tidal tail. It exhibits a strikingly strong
velocity gradient the origin of which could be multifold, including
rotation.  The TDG is currently experiencing a strong starburst.

We have a wealth of spectrophotometric data which we still have to
analyze in detail,in particular to derive reddening, star-formation rate,
and metal abundance. Several more TDG candidates with kinematical
signatures are visible on 2D spectra (see Tab.~\ref{tab:sample}); we
can therefore expect to find a few more true TDGs in our sample.

Comparison with photometry and spectra from evolutionary synthesis
models (see Weilbacher \& Fritze-v.Alvensleben, this volume) will
allow to reasonably constrain the burst parameters and provide clues
to the future evolution of our TDG candidates.

\vspace{.3cm}
{\small {\bf Acknowledgement}. PMW is partially supported by the Deutsche
  Forschungsgemeinschaft (DFG Grant FR 916/6-1).}

\renewcommand{\refname}{{\small\bf References}}
\itemsep=0pt \parsep=0pt \parskip=0pt \labelsep=0pt
{\small
  \bibsep=0pt
  \bibliography{../../../PmW}
}

\end{document}